\documentclass[pra,twocolumn, showpacs]{revtex4}
\usepackage{epsfig}

\begin{document}
\title{Witnessing Macroscopic Entanglement in a Staggered Magnetic Field}
\author{Jenny Hide$^1$, Wonmin Son$^1$, Ian Lawrie$^1$, Vlatko Vedral$^{1,2}$}
\affiliation{School of Physics and Astronomy, EC Stoner Building,
University of Leeds, Leeds, LS2 9JT, UK$^1$\\
Quantum Information Technology Lab, Department of Physics, National
University of Singapore, Singapore 117542 $^2$}

\date{\today}

\begin{abstract}
We investigate macroscopic entanglement in an infinite XX 
spin-$\frac{1}{2}$ chain with staggered magnetic field, $B_l=B+e^{-i
\pi l}b$. Using single-site entropy and by constructing an
entanglement witness, we search for the existence of entanglement
when the system is at absolute zero, as well as in thermal
equilibrium. Although the role of the alternating magnetic field $b$
is, in general, to suppress entanglement as do $B$ and $T$, we find
that when $T=0$, introducing $b$ allows the existence of
entanglement even when the uniform magnetic field $B$ is
arbitrarily large. We find that the region and the amount of entanglement
in the spin chain can be enhanced by a staggered magnetic field.
\end{abstract}

\maketitle


Quantum entanglement is a fundamental aspect of quantum physics. It
demonstrates the non-local nature of the theory in that an entangled
system contains correlations that cannot be
described by its subsystems alone. Instead these
quantum correlations are attributed to the overall system \cite{EPR}.
Further, entanglement is an important resource in quantum
information and computation. In particular, solid state quantum
computation has become a topic of much research and several
proposals for physical implementation have been investigated.  The
Heisenberg interaction is the model used in many physical
applications of quantum computation, for example, quantum dots
\cite{dots} and cavity QED \cite{Imamoglu99}. It has also been shown
that the Heisenberg interaction can be used to implement any circuit
required by a quantum computer \cite{divi}. Therefore, entanglement
in one-dimensional spin chains has been the subject of much
interest. This entanglement has been studied both in the case of a
finite spin chain \cite{wang, arnesen} and in the thermodynamic
limit \cite{vedral} where the length of the spin chain becomes
infinite.

\emph{Macroscopic} entanglement is a more recent concept. It
demonstrates that non-local correlations persist even in the
thermodynamic limit. This type of entanglement can be detected by
measuring macroscopic quantities such as internal energy and
magnetic susceptibility \cite{mag_sus} as it has been proven that
such quantities can be used as entanglement witnesses. It has been
shown experimentally \cite{ghosh, vedral_nat}, that the behaviour of
observable macroscopic quantities such as magnetic susceptibility
depends, most significantly at low temperatures, on entanglement.
This demonstrates that entanglement is vital in the explanation of
how macroscopic materials behave. Macroscopic entanglement in a
Heisenberg spin chain has been studied previously \cite{vedral} only
for a uniform magnetic field. The Hamiltonian of this chain is used 
to construct an entanglement witness \cite{dowl1, toth1, wu1} which 
shows that entanglement disappears for high uniform magnetic field 
just as it does for high temperature.

In real systems, the magnetic field need not be the same at each
site in the chain.  In solid state systems, there exists a
possibility that an inhomogeneous Zeeman coupling could induce a
non-uniform magnetic field. Moreover, an experimental system is
likely to contain magnetic impurities. Copper Benzoate \cite{Cu_B}, 
\cite{Cu_B2} is a practical example of a system in a
non-uniform magnetic field. In this case, the alternating field is in 
a direction perpendicular to the uniform field. Alternatively, such
impurities could be introduced artificially. Therefore, the
possibility that such a field could affect entanglement, whether to
reduce or increase it, is an important subject to investigate. In
reality, systems have a finite temperature so the thermal case must
be considered. Hence, in this paper, we discuss the effect of a
site dependent magnetic field on thermal macroscopic entanglement in
a 1-D infinite spin-$\frac{1}{2}$ chain. We also consider the zero
temperature case. Interestingly, we show that an alternating magnetic
field can compensate for the effect of a uniform magnetic field at
$T=0$.

The Hamiltonian considered is
\begin{equation}
H=-\sum_{l} \left[ \frac{J}{2}\left(\sigma_l^x
\otimes\sigma_{l+1}^x+\sigma_l^y \otimes
\sigma_{l+1}^y\right)+B_l^{}\sigma_l^z \right]
\label{eq:hamiltonian1}
\end{equation}
where $J$ is the coupling strength between sites, and 
$B_l^{}=B+e^{-i\pi l}b$ is the site dependent magnetic field.  
Although such a field 
is not likely to occur in nature, our work allows 
us to investigate how a non-uniform magnetic field affects 
entanglement in a model which is analytically solvable. A similar 
Hamiltonian with cyclic boundary conditions has previously been 
diagonalized \cite{suzuki} using a method first set out by S.
Katsura \cite{katsura, kat2}.  In our discussions of a finite
$N$-spin chain, we consider the case of open boundary conditions
with $N$ even. In fact, these constraints are no longer relevant in
the thermodynamic limit and hence our conclusion is unchanged if,
for example, the cyclic model is used.

To identify entanglement in this system, we use an entanglement
witness, i.e. an operator whose expectation value is bounded for any
separable state. The power of our witness is such that we can
identify the existence of entanglement even for a thermal system
which is a mixed state in general.   Alternatively, single-site
entropy can be taken as evidence of entanglement when $T=0$ since
the total system is in a pure state. The purity of the single-site
density matrix shows that the entanglement witness is not optimal
at absolute zero.  Thus we find that although the alternating magnetic
field, $b$, acts in general to suppress entanglement similarly to
$B$ and $T$, at zero temperature, increasing $b$ allows the system
to be entangled for arbitrarily large $B$.  Hence the effect of the
staggered field at $T=0$ is to increase both the amount and the region of
entanglement.

{\it Entanglement} - A system is said to be entangled when its
density matrix cannot be written as a convex sum of product states. For 
a pure state, dividing the system into two subsystems $A$ and $B$ allows 
the von Neumann entropy to be used as a measure of entanglement.  If we
trace section $B$ out of the density matrix to find $\rho_A$, the
von Neumann entropy, $S(\rho_A)=-\mbox{Tr}(\rho_A\log_2\rho_A)$, can
be calculated. $S(\rho_A)=0$ corresponds to a separable state while
when $S(\rho_A)=1$, the system is maximally entangled.  In the case of
a mixed state, there is no unique measure of entanglement for a
multipartite system. However, we can construct an entanglement witness.

An entanglement witness \cite{horod} is an operator whose
expectation value for any separable state is bounded by a value
corresponding to a hyperplane in the space of density matrices. An
entanglement witness is only a sufficient condition for the
existence of entanglement. Hence failure of the witness to detect
entanglement does not necessarily mean the system is separable.
Though witnesses simply detect rather than give a measure of
entanglement, they have significant advantages over other methods.
For example, they naturally incorporate temperature and many witnesses, 
such as magnetic susceptibility, can be experimentally measured
\cite{mag_sus}.

{\it The partition function and entanglement witness} - Many
thermodynamic variables can be derived from the Helmholtz free
energy, $F=-T\ln Z$, where $Z$ is the partition function. As 
$\partial F/\partial X =\left\langle\frac{\partial H}{\partial
X}\right\rangle$, we see that when $X=B$, we obtain the magnetization 
$M=\sum_{l}\langle\sigma_l^z\rangle=-\partial F/\partial B$. In 
particular, we define the entanglement witness
\begin{equation}
W=\frac{2}{\beta N}\frac{\partial \ln Z}{\partial J}
=\frac{1}{N}\sum_{l} \Big(\langle\sigma_l^x\sigma_{l+1}^x\rangle+
\langle\sigma_l^y\sigma_{l+1}^y\rangle\Big) \label{eq:entW}
\end{equation}
where $\beta=1/T$.  Our witness, W,  identifies a larger entangled region
than witnesses found previously \cite{vedral}.  In a
separable state, it satisfies the bound $\vert W\vert\le 1$, which
can be shown as follows. With $\rho=\sum_i p_i
\rho_1^i\otimes\rho_2^i\otimes\cdot\cdot\cdot\otimes\rho_N^i$, we
have $|\langle\sigma_l^x\sigma_{l+1}^x\rangle+
\langle\sigma_l^y\sigma_{l+1}^y\rangle|=|\
\langle\sigma_l^x\rangle\langle\sigma_{l+1}^x\rangle+
\langle\sigma_l^y\rangle\langle\sigma_{l+1}^y\rangle| \leq
\sqrt{\langle\sigma_l^x\rangle^2+\langle\sigma_l^y\rangle^2}
\sqrt{\langle\sigma_{l+1}^x\rangle^2+\langle\sigma_{l+1}^y\rangle^2}\leq
1$ for any $l$. The upper bound for the inequality is found by using
the Cauchy-Schwarz inequality and the condition that for any state,
$\langle\sigma_l^x\rangle^2+\langle\sigma_l^y\rangle^2+
\langle\sigma_l^z\rangle^2\leq 1$. Thus, any state that violates the
inequality $\vert W\vert\le 1$ is entangled.

\begin{figure}[t]
\begin{center}
\centerline{
\includegraphics[width=2.6in]{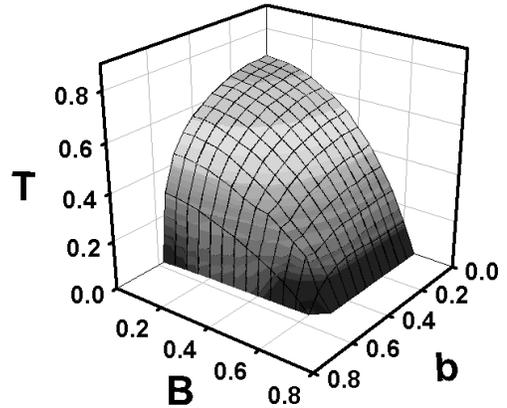}   }
\end{center}
\caption{The surface $\vert W\vert=1$ in the space of magnetic
field, $B$, temperature, $T$, and alternating magnetic field, $b$.
The region between the surface and the axes is entangled.} \label{fig1}
\end{figure}

{\it Diagonalization of Hamiltonian} - In order to find the partition
function of the system, we must diagonalize the Hamiltonian,
Eq. (\ref{eq:hamiltonian1}) and find its energy eigenvalues.
The open ended Hamiltonian can be exactly diagonalized in a standard
way via several steps: a Jordan-Wigner tansformation, a Fourier
transformation and finally a Bogoliubov transformation. The Jordan-Wigner
transformation,
\begin{equation}
a_l=\prod_{m=1}^{l-1} \sigma_m^z \otimes\frac{ \left(\sigma_l^x
+i\sigma_l^y \right)}{2}, \label{eq:JW_trans}
\end{equation}
maps the Pauli spin operators into fermionic annihilation and
creation operators $a_l$ and $a_l^{\dagger}$.  These satisfy the
anti-commutation relations $\{a_l, a_k\}=0$ and $\{a_l,
a_k^{\dagger}\}=\delta_{l,k}$. Preserving the anti-commutation
relations, the operators can now be transformed unitarily using a
Fourier transformation, $a_l = \sqrt{\frac{2}{N+1}}\sum_{k=1}^N d_k
\sin(\frac{\pi kl }{N+1})$ and by a Bogoliubov transformation,
\begin{eqnarray}
d_k & = & \alpha_k \cos \theta_k +\beta_k \sin \theta_k \\ \nonumber
d_{N+1-k} & = & \beta_k \cos \theta_k -\alpha_k \sin \theta_k.
\label{eq:can_trans}
\end{eqnarray}
Setting $\tan 2\theta_k =b/[J \cos(\pi k /(N+1))]$ eliminates the
off-diagonal terms leaving the Hamiltonian in diagonal form
\begin{equation}
H=\sum_{k=1}^{N/2}\left( \lambda_k^{+}\alpha_{k}^\dagger \alpha_{k}
+\lambda_k^{-}\beta_{k}^\dagger \beta_{k}-2 B{\bf 1} \right)
\label{eq:diagon}
\end{equation}
where $\lambda_k^{\pm}  = 2B\pm 2\sqrt{J^2\cos^2(\pi k/(N+1))+b^2}$.
The operators $\alpha_k$ and $\beta_k$ satisfy the anti-commutation
relations
$\{\alpha_k,\alpha_l^{\dagger}\}=\{\beta_k,\beta_l^{\dagger}\}=\delta_{k,l}$
and $\{\alpha_k,\beta_l\}=\{\alpha_k,\beta_l^\dag\}=0$.
Using the eigenvalues of the Hamiltonian, we find that the partition
function can be written $Z=\prod_{k=1}^{N/2}2\cosh\left(\beta
\lambda_k^+/2\right)2\cosh\left(\beta \lambda_k^-/2\right)$. In the
thermodynamic limit, $N\rightarrow \infty$, we can treat $\omega=
\pi k/N$ as a continuous variable, to find
\begin{equation} \ln Z=\frac{N}{\pi}\int_0^{\pi/2}
d\omega \ln\left[4\cosh\left(\frac{\beta
\lambda^+_\omega}{2}\right)\cosh\left(\frac{\beta
\lambda^-_\omega}{2}\right)\right] \label{lnZ}
\end{equation}
where $\lambda_{\omega}^{\pm}  = 2B\pm
2\sqrt{J^2\cos^2\omega+b^2}$. We can now use Eq. (\ref{eq:entW}) to
calculate the entanglement witness for our system.

The region of entanglement detected by our witness has been plotted in
Fig. (\ref{fig1}). The figure shows the region of uniform magnetic field,
$B$, temperature, $T$, and alternating magnetic field, $b$, within
which we always find entanglement.  At fixed values of $b$, the
entangled region of the $T-B$ plane shrinks as $b$ increases until a 
critical value is reached above which entanglement is
no longer detected. Consider now a plane perpendicular to the $T$
axis.  At zero temperature we see that until the critical value
$b_c=0.56$, increasing $b$ has no effect on the value that the
uniform magnetic field can take with the system remaining in an
entangled state. Above $b_c$, our witness detects no entanglement.
However, we later show that this witness is not optimal at zero
temperature.

Our witness shows entanglement behaving as we would expect physically.
A high temperature causes the system to become mixed.  Thus no
quantum correlations can survive and the system is separable.
Moreover, the effect of the magnetic fields is understandable as local
operations which enhance classical correlations. When the uniform
magnetic field becomes large, spins tend to line up in a
direction parallel to that field. This clearly decreases quantum
correlations in the system. Using the same reasoning, if the alternating
magnetic field is large, the spins tend to anti-align which
is also a product state. Hence, as Fig. (\ref{fig1}) shows, all of
these parameters cause the system to become separable if they are
large enough.  Interestingly, we find a counter example of this
expected behaviour and identify a region where the system is
entangled even in a large magnetic field.  We discuss this
in the following section.

{\it Single-site entropy} - When $T=0$, the total system is in a
pure state, i.e. the ground state. Thus, if the density matrix of a
single spin is in a mixed state, the particle must be entangled with
the rest of the spins. This can be quantified using the entropy of a
single spin.  The $l$-th spin density matrix, $\rho_l=\frac{1}{2}
\sum_{i\in\{1,x,y,z\}} \sigma_l^i \langle \sigma_l^i \rangle$, can
be obtained from the total density matrix
$\rho=e^{-\beta\hat{H}}/Z$. Moreover, we find that the single-site
density matrix is readily diagonalized since
$\langle\sigma_l^x\rangle=\langle\sigma_l^y\rangle=0$. This follows
from the fact that $\sigma_l^x$ and $\sigma_l^y$ are linear
combinations of the fermion operators $\alpha_k$, $\alpha_k^\dag$,
$\beta_k$, and $\beta_k^\dag$, all of which have zero expectation
values.

In the thermodynamic limit $N\to\infty$, the system is translationally
invariant for all odd sites and for all even sites.  Hence the
single-site  magnetization $\langle\sigma_l^z\rangle$ can
be obtained from the total
magnetization $M$ and the total staggered magnetization $M_{\mathrm
s}$.  In the limit of zero temperature, these are given by
\begin{eqnarray}
\label{eq:single-spin} M&=&\sum_l\langle \sigma_l^{z}\rangle
=\frac{1}{\beta}\frac{\partial}{\partial B} \ln Z
=N\left(1-\frac{2\Omega}{\pi}\right)\\
M_{\mathrm s}&=&\sum_l(-1)^l\langle \sigma_l^{z}\rangle
=\frac{1}{\beta}\frac{\partial}{\partial b} \ln Z
=\frac{N}{\pi}\int_0^{\Omega} d\omega f(\omega) \nonumber
\end{eqnarray}
where $f(\omega)=2 b/\sqrt{J^2\cos^2\omega+b^2}$ and
$$
\Omega = \left\{ \begin{array}{lcl}
\pi/2 & \mbox{for}& B< b \\
\cos^{-1}\Big(\sqrt{B^2-b^2}/J\Big) & \mbox{for}
& b\leq B\leq\sqrt{J^2+b^2} \\
0 & \mbox{for} & B>\sqrt{J^2+b^2}
\end{array}\right.
$$
To obtain these results, we have used
$\lim_{\beta\rightarrow\infty}\tanh(\beta x)= x/|x|$. By virtue of
translational invariance,
$\langle \sigma_l^{z}\rangle$, is the same for all even sites and for all
odd sites.  Thus defining
$\langle\sigma^z_l\rangle=\mbox{Tr}(\sigma^z_l \rho)= 2 \eta_l -1$,
such that the single-site density matrix is
$\rho_l=\mbox{diag}(\eta_l, 1-\eta_l)$, we have
\begin{equation}
\eta_{l}=\frac{1}{2}\Big\{1+\frac{1}{N} \left[M+(-1)^l
M_\mathrm{s}\right]\Big\}.
\end{equation}
From this, we can obtain the entropy of the $l$-th spin,
$S=-\eta_l\log_2\eta_l-(1-\eta_l)\log_2(1-\eta_l)$.

\begin{figure}[t]
\begin{center}
\centerline{
\includegraphics[width=1.6in]{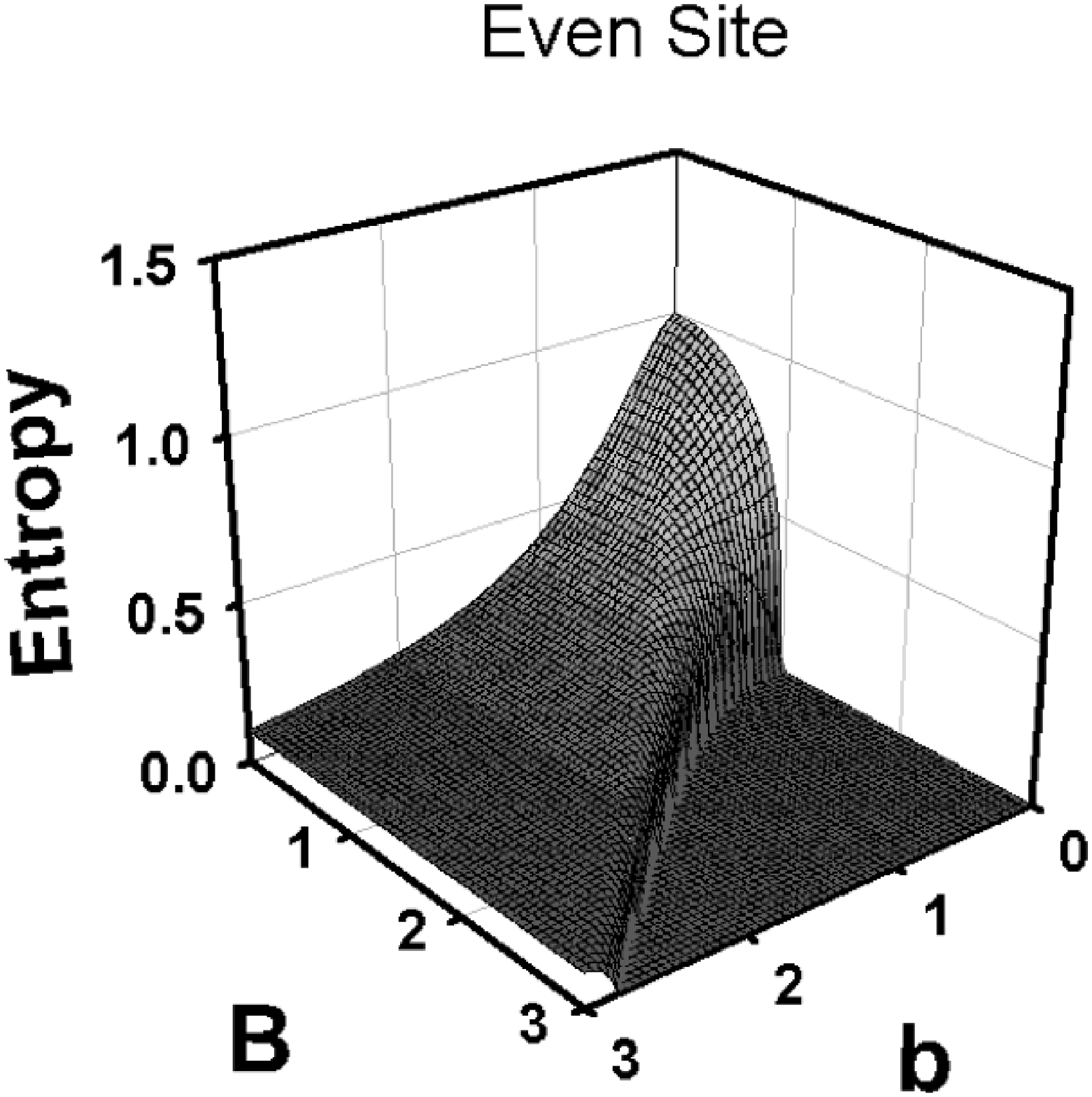}
\includegraphics[width=1.7in]{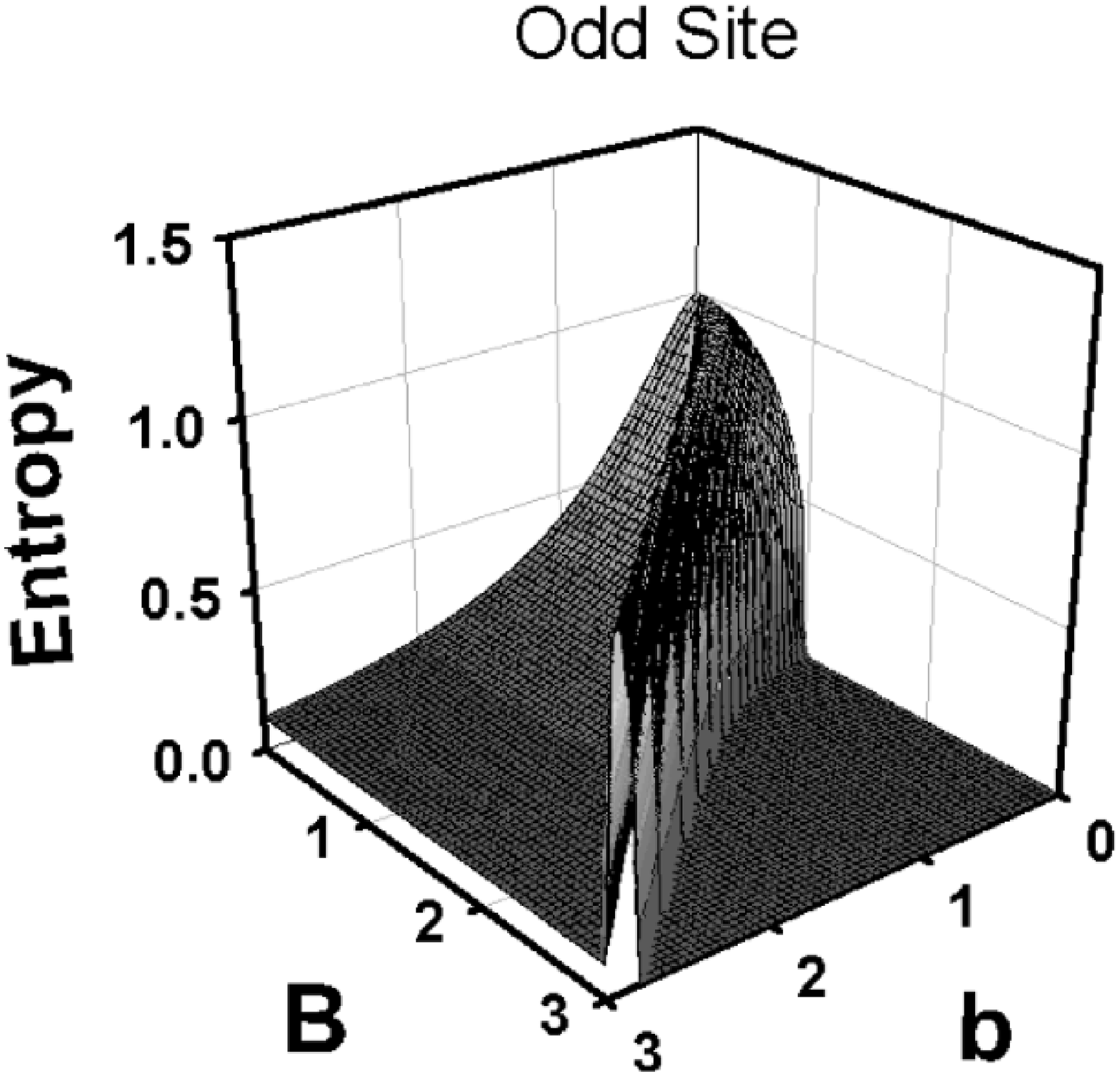} }
\end{center}
\caption{We plot the entropy for an even site spin and an odd site
spin respectively when $T=0$ and $J=1$. Excluding the region
$B>\sqrt{1+b^2}$, the single-site entropy is non zero and each spin
is entangled with the rest of the system. The odd site entropy shows
a maximum when $B=b+\varepsilon$.} \label{fig2}
\end{figure}

\begin{figure}[t]
\begin{center}
\centerline{
\includegraphics[width=2.6in]{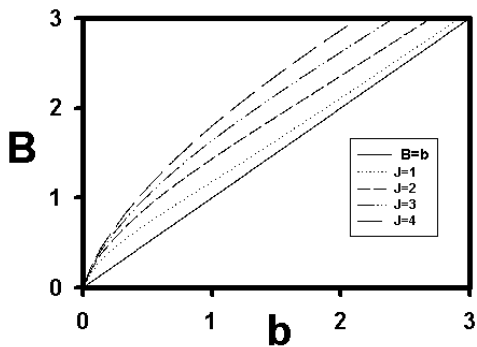} }
\end{center}
\caption{Plot of the maximum entropy $S_l=1$ for odd site spins
with different values of $J$. As the value of $J$ increases, this 
maximum entropy moves further away from the line $B=b$.
} \label{fig3}
\end{figure}

The single-site entropy for odd and even sites is plotted in Fig.
(\ref{fig2}) with $J=1$. In both cases, we find entanglement for
various values of $B$ and $b$ even when the magnetic fields are
large. The entropy, and therefore entanglement between
each spin and the remainder of chain, is non-zero everywhere except when
$B>\sqrt{J^2+b^2}$. Hence entanglement exists when the coupling
strength between nearest neighbour spins, $J$, is more than
$\sqrt{B^2-b^2}$.  We note that this corresponds to when the square
of the interaction strength is greater than the product of the total
magnetic field on two adjacent sites. In addition, we observe from
Fig. (\ref{fig2}) that the maximum single-site entropy occurs when
both magnetic fields $B$ and $b$ are zero. Introducing the
magnetic fields reduces the amount of entanglement in the system 
except along the peak in the odd site entropy. We find from the
maximum entropy, $S_l=1$, that the maximum entanglement occurs in
the region $b\leq B \leq \sqrt{J^2 +b^2}$ when $\Omega-\pi/2=(-1)^l
\int_0^{\Omega}b/\sqrt{J^2 \cos^2 \omega +b^2} d\omega$ is
satisfied. This corresponds to when the magnetization, $M$, is 
equal to the staggered magnetization, $M_\mathrm{s}$.
For even sites, there is only one solution to this 
at $B=b=0$. For odd sites however, this is satisfied for
any finite uniform magnetic field $B$ at $B=b+\varepsilon$ where
$\varepsilon$ is a positive value.  The solutions in Fig. (\ref{fig3}) 
correspond to the peak in the odd site entropy and so occur
within the range $b\leq B \leq \sqrt{J^2+b^2}$. In general, 
$\varepsilon$ becomes larger as $J$ increases as shown in 
Fig. (\ref{fig3}), and becomes smaller as $B$ and $b$ increase.  
We note that the
peak does not occur at $B=b$. As $B$ and/or $b$ tend to infinity, 
the amount of entanglement in the system tends to zero.  Further, 
as the magnetic fields increase, the curves in Fig. (\ref{fig3}) 
tend to the $B=b$ line.  At $B=b$, 
the system is no longer maximally entangled. At this point, 
we find from the Hamiltonian that odd sites have zero magnetic 
field, and even sites have field strength $2B$. Hence at the 
peak, odd site spins have a small magnetic field, $\varepsilon$, 
while in comparison, even site spins have a large field, 
$2B-\varepsilon$. The implications of this peak are that even 
in the limit of large (though not infinite) uniform field $B$, an 
odd site can be maximally entangled with the
rest of the system if an appropriate alternating magnetic field, $b$
is introduced. In practice, fine-tuning $b$ to achieve maximal 
entanglement may be difficult. However, for any $B$, sufficiently 
increasing $b$ (when $b>\sqrt{B^2-J^2}$) will create entanglement in 
the chain. Hence a staggered magnetic field can enhance the
amount of entanglement present in the system.

Although the single-site entanglement relates only to the zero
temperature case, changing the temperature by a small amount should
not change its behaviour.  Hence for very low temperatures, we see
that the entanglement witness is not optimal.

{\it Conclusions} - Our best estimate for finite temperature
entanglement is the witness which shows $b$ reduces the region of
entanglement in the chain. That is, a staggered magnetic field 
reduces the entangled region.  If this behaviour is true even for an
optimal finite temperature witness, these results have consequences
for larger scale quantum computation in solid state systems.  As
inhomogeneities in the magnetic field exist naturally in the Zeeman
coupling between atoms, the region of entanglement is naturally
decreased compared to when $b=0$.  Quantum computation relies on
entanglement so as introducing $b$ reduces both the temperature and
uniform magnetic field at which entanglement persists, our result
shows it may be more difficult than previously thought to construct
useful quantum computers using one-dimensional systems. 

Our entanglement witness is invaluable as by applying it to our
system, it allows us to see how temperature affects the entanglement
in the spin chain. However, the witness does not tell us how the
entanglement actually behaves in the presence of $B$, $T$ or $b$ as
it does not detect all entanglement in the system. Conversely, the
single-site entropy shows us exactly how the entanglement is
affected by the magnetic fields at zero temperature, although it is
unknown how to extend this entropy to a finite temperature.  Using 
this entropy, we have shown that in the thermodynamic limit, 
a staggered magnetic field enhances both the region and the amount 
of entanglement in our spin chain. Hence, both the witness and the 
entropy are essential in characterizing the entanglement in
the system.  Further, the region of entanglement identified by 
the witness is consistent with that of the single-site 
entropy.  If the behaviour of the entanglement as shown by the 
entropy persists at higher temperatures, we may be able to 
counteract any Zeeman coupling by applying an appropriate magnetic 
field, hence maximizing entanglement for odd sites.  This will be 
an interesting topic for future research.

{\it Acknowledgement} - V.V. \& J.H. acknowledge the EPSRC for financial 
support.


\end{document}